\documentclass[aps,pra,superscriptaddress,reprint,showpacs,nofootinbib]%
{revtex4-1}
\usepackage{amsfonts,amsmath,graphicx,xspace}

\newcommand{\BB}{BB84\xspace}
\newcommand{\CSS}{CSS\xspace}
\newcommand{\QKD}{QKD\xspace}
\newcommand{\EDQKD}{ED-QKD\xspace}
\newcommand{\PMQKD}{PM-QKD\xspace}
\newcommand{\Chauzerofive}{Chau05\xspace}
\newcommand{\Chauonefive}{Chau15\xspace}
\newcommand{\SchemeA}{Scheme~A\xspace}
\newcommand{\SchemeB}{Scheme~B\xspace}
\newcommand{\SchemeC}{Scheme~C\xspace}
\newcommand{\RRDPS}{RRDPS\xspace}

\DeclareMathOperator{\Tr}{Tr}

\begin{document}
\title{Experimentally Feasible Quantum-Key-Distribution Scheme Using
 Qubit-Like Qudits And Its Comparison With Existing Qubit- and Qudit-Based
 Protocols}
\author{H. F. Chau}
\thanks{Corresponding author, email: \texttt{hfchau@hku.hk}}
\affiliation{Department of Physics, University of Hong Kong, Pokfulam Road,
 Hong Kong}
\affiliation{Center of Theoretical and Computational Physics, University of
 Hong Kong, Pokfulam Road, Hong Kong}
\author{Qinan Wang}
\affiliation{Department of Physics, University of Hong Kong, Pokfulam Road,
 Hong Kong}
\author{Cardythy Wong}
\affiliation{Department of Physics, University of Hong Kong, Pokfulam Road,
 Hong Kong}
\date{\today}
\begin{abstract}
 Recently, Chau~\cite{Chau15} introduced an experimentally feasible qudit-based
 quantum-key-distribution (\QKD) scheme.
 In that scheme, one bit of information is phase encoded in the prepared state
 in a $2^n$-dimensional Hilbert space in the form
 $(|i\rangle\pm|j\rangle)/\sqrt{2}$ with $n\ge 2$.
 For each qudit prepared and measured in the same two-dimensional Hilbert
 subspace, one bit of raw secret key is obtained in the absence of transmission
 error.
 Here we show that by modifying the basis announcement procedure, the same
 experimental setup can generate $n$ bits of raw key for each qudit prepared
 and measured in the same basis in the noiseless situation.
 The reason is that in addition to the phase information, each qudit also
 carries information on the Hilbert subspace used.
 The additional $(n-1)$ bits of raw key comes from a clever utilization of this
 extra piece of information.
 We prove the unconditional security of this modified protocol and compare its
 performance with other existing provably secure qubit- and qudit-based
 protocols on market in the one-way classical communication setting.
 Interestingly, we find that for the case of $n=2$, the secret key rate of this
 modified protocol using non-degenerate random quantum code to perform
 one-way entanglement distillation is equal to that of the six-state scheme.
\end{abstract}


\maketitle

\section{Introduction}
\label{Sec:Intro}
 Prepare-and-measure-based quantum-key-distribution (\PMQKD) protocol is a
 class of practical schemes in which the sender Alice prepares a quantum state
 and sends it through an insecure channel to the receiver Bob, who measures the
 received state so as to establish a shared raw key.
 Then, they apply classical post-processing to the raw key to distill a secure
 final
 key~\cite{*[{See, for example, }] [{ and references cited therein.}] RMP09}.
 While early \PMQKD protocols such as the well-known \BB scheme~\cite{BB84}
 use unentangled qubits as quantum information carriers, various authors
 proposed using qudits
 instead~\cite{Chau15,RMP09,BP00,BT00,CBKG02,Chau05,SYK14}.
 Generally speaking, qudit-based schemes are more error tolerant than
 qubit-based ones.
 However, qudit-based schemes are generally very hard to implement in practice
 partly because of the difficulty in preparing a general qudit state with high
 fidelity.
 Two notable exceptions are the recently proposed round-robin
 differential-phase-shift (\RRDPS) scheme~\cite{SYK14} and the so-called
 \Chauonefive scheme~\cite{Chau15}.

 Recall that for the \Chauonefive scheme, Alice randomly picks two distinct
 elements $i,j$ from the Galois field $GF(2^n)$ with $n\ge 2$ and prepares a
 state in the form $(|i\rangle\pm|j\rangle)/\sqrt{2}$, where $\{ |i\rangle
 \colon i\in GF(2^n) \}$ is an orthonormal basis of the $2^n$-dimensional
 Hilbert space.
 After receiving this state from Alice, Bob randomly picks two distinct
 elements $i',j' \in GF(2^n)$ and projectively measures the state along $\{
 (|i'\rangle\pm|j'\rangle)/\sqrt{2}$.
 The \Chauonefive scheme is experimentally feasible because in the time bin
 representation, the preparation and measurement procedures are almost
 identical to those for diagonal basis qubits~\cite{Chau15}.
 In this regard, we call these preparation and measurement states qubit-like.
 The security of the \Chauonefive scheme originates from the fact that for $n
 \ge 2$, the values of $i,j,i',j'$ used, and hence the Hilbert subspace picked
 during the preparation and measurement of these qubit-like states, are
 withheld from the eavesdropper Eve until after Bob's measurement.
 Clearly, Alice and Bob should get a shared raw bit of key encoded in the phase
 of the prepared qudit should $\{i,j\} = \{i',j'\}$.
 In other words, the \Chauonefive scheme is able to generate one bit of raw
 secret key per successful transfer of each $2^n$-dimensional qudit provided
 that it is prepared and measured in the same Hilbert subspace~\cite{Chau15}.

 Here we show that the \Chauonefive scheme can be modified so that the number
 of raw secret bits generated per each such successful qudit transfer can be
 increased from $1$ to $n$.
 We do it by replacing the announcement procedures for the Hilbert subspaces
 used to the preparation and measurement bases.
 In this way, the $(n-1)$ classical bits used to describe the Hilbert subspace
 information of each prepare and measure qubit-like qudit state, which is also
 withheld from Eve until Bob's measurement, can then be used to generate part
 of the raw key.
 More importantly, there is no need to change the hardware setup of the
 \Chauonefive scheme in this modification.

 In Sec.~\ref{Sec:Scheme}, we first introduce an
 entanglement-distillation-based quantum key distribution (\EDQKD) protocol
 known as \SchemeA.
 Then, we use Shor-Preskill argument~\cite{RMP09,SP00} to show that \SchemeA
 can be reduced to two equally secure \PMQKD protocols known as Schemes~B
 and~C.
 In particular, the state preparation and measurement procedures in \SchemeC
 are identical to that of the \Chauonefive scheme.
 We then prove the unconditional security of \SchemeA and give key rate
 formulas under one-way entanglement distillation for Schemes~A--C in
 Sec.~\ref{Sec:Security}.
 We also compare the performance of our schemes to various provably secure
 qubit- and qudit-based \PMQKD schemes in the literature using one-way
 entanglement distillation in Sec.~\ref{Sec:Security}.
 In particular, we find that for the case of $n=2$, the secret key rate of
 \SchemeB is equal to that of the six-state scheme~\cite{Bruss98} when both use
 non-degenerate random quantum codes to perform one-way entanglement
 distillation.
 Finally, we briefly discuss the experimental feasibility of Schemes~B and~C.

\section{The Modified Schemes}
\label{Sec:Scheme}

\subsection{The Entanglement-Distillation-Based Scheme Known As \SchemeA}
\label{Subsec:EDQKD}
 Let $N \equiv 2^n$ with $n\ge 2$ and consider the following \EDQKD
 scheme known as \SchemeA.\footnote{From now on, we use the convention that a
 variable in Roman, square-bracketed, overbarred and Greek alphabet are in
 $GF(N)$, $GF(N)/GF(2)$, $GF(2)$ and $GF(N)^*$, respectively.}
 (The description below extensively uses a lot of finite field arithmetic.
 Readers may consult Ref.~\cite{FF} for an introduction.)

\par\medskip
{\bf The Modified \EDQKD Scheme (\SchemeA).}
\begin{enumerate}
 \item Alice secretly and randomly picks $[a]\in GF(N)/GF(2)$ and $\lambda\in
  GF(N)^* \equiv GF(N) \setminus \{0\}$.
  She prepares the state $\sum_{\bar{i}\in GF(2)} |\bar{i}+[a]\rangle_\text{A}
  \otimes |\bar{i}+[a]\rangle_\text{B} / \sqrt{2}$, where all arithmetic in the
  ket state are performance in the finite field $GF(N)$.
  She applies the linear transformation $L_\lambda|i\rangle \mapsto |\lambda i
  \rangle$ for all $i\in GF(N)$ to the second qudit before sending it through
  an insecure quantum channel to Bob.
  \label{Ent:prepare}
 \item Upon reception of the state from Alice, Bob secretly and randomly picks
  $\lambda' \in GF(N)$ and applies the linear transformation
  $L_{\lambda'}^{-1}$ to his received state.
  \label{Ent:transform}
 \item Alice and Bob jot down the joint measurement result along the basis
  \begin{align}
   {\mathcal B} &= \{ \frac{1}{\sqrt{2}} \sum_{\bar{i}\in GF(2)} (-1)^{\bar{i}
    \bar{c}} |\bar{i}+[a]\rangle_\text{A} \otimes |\bar{i}+[a]+b
    \rangle_\text{B} \nonumber \\
   & \qquad \quad \colon [a]\in GF(N)/GF(2), b\in GF(N), \nonumber \\
   & \qquad \quad \bar{c}\in GF(2) \} \nonumber \\
   &\equiv \{ |\Phi_{[a],b,\bar{c}}\rangle \}
   \label{E:Bcal_def}
  \end{align}
  to their shared quantum state.
  Then, they publicly announce the values of $\lambda,\lambda'$ used and keep
  the state only if $\lambda=\lambda'$.
  They repeat steps~\ref{Ent:prepare}--\ref{Ent:measure} until they have enough
  shared pairs.
  \label{Ent:measure}
 \item Alice and Bob pick a random sample from their remaining measured states
  and reveal the values of $b$ and $\bar{c}$ obtained for each of the selected
  states for various $\lambda$'s and $[a]$'s to estimate the error rate of the
  channel.
  Specifically, let $\tilde{e}_{b\bar{c}}$ be the probability that Alice
  prepares the state $|\Phi_{[a],0,0}\rangle$ and that the resultant state
  measured by Alice and Bob is $|\Phi_{[a],b,\bar{c}}\rangle$ for some $[a]\in
  GF(N)/GF(2)$.
  Then, by revealing the values of $b$ and $\bar{c}$ from a random sample of
  those shared states to which Alice and Bob have applied $I\otimes L_\lambda$
  and $I\otimes L_\lambda^{-1}$, they obtain an estimate of the value of
  $\tilde{e}_{\lambda [b],\bar{c}} + \tilde{e}_{\lambda ([b]+1),\bar{c}}$ for
  all $[b]\in GF(N)/GF(2)$ and $\bar{c}\in GF(2)$.
  They proceed only if the error rate is sufficiently small.
  (We shall discuss the smallness criterion later on in the unconditional
  security proof in Sec.~\ref{Sec:Security}.)
  \label{Ent:error-rate_test}
 \item Alice and Bob apply one- or two-way entanglement distillation similar to
  the ones used in Refs.~\cite{Chau15,SP00,L01,GL03,Chau05} to the remaining
  states to distill out almost perfect EPR-like states each in the form
  $|\Phi_{[a],0,0}\rangle$.
  For instance, they apply a Calderbank-Shor-Steane (\CSS) quantum
  error-correcting code~\cite{CS96,Steane96,Rains99,AK01} that could correct
  the measured spin-flip and phase errors of the channel in
  step~\ref{Ent:error-rate_test}.
  (Note that such a \CSS code is constructed using a classical $N$-ary code
  $C_1$ and a classical binary code $C_2$ obeying $\{0\} \subset C_2 \subset
  C_1$ via the standard \CSS construction.
  This is possible for a binary code can be regarded as an $N$-ary code by
  extending the linear coding space over the field $GF(2)$ to the linear space
  over the field $GF(N)$.
  In fact, we may extend the dual code of the binary code $C_2$ to an $N$-ary
  code with the same minimum distance using the same trick.
  In this way, $C_1$ and $C_2$ can be used to correct spin-flip and phase
  errors in this noisy and insecure channel, respectively.
  More importantly, the choice of $C_1$ could depend on the error syndrome
  measurement results of the code $C_2$ just like the one used by Lo in
  Ref.~\cite{L01}.)
  \label{Ent:priv_ampl}
 \item Finally, Alice and Bob separately measure each of their share of the
  almost perfect EPR-like states along the basis $B_1$, where
  \begin{align}
   B_\lambda &= \{ \frac{1}{\sqrt{2}} \sum_{\bar{i}\in GF(2)} (-1)^{\bar{i}
    \bar{c}} |\lambda(\bar{i}+[a])\rangle \colon \nonumber \\
   & \qquad \quad [a]\in GF(N)/GF(2), \bar{c}\in GF(2) \}
   \label{E:basis_def}
  \end{align}
  for all $\lambda\in GF(N)^*$.
  In this way, they obtain $n$~bits of shared secret key per EPR-like state
  measured --- 1~bit comes from the phase information $\bar{c}$ and
  $(n-1)$~bits come from the value of $[a]\in GF(N)/GF(2)$.
  \label{Ent:final_key}
\end{enumerate}

 We remark that in the absence of noise and Eve, Alice and Bob should get $b =
 \bar{c} = 0$ for each pair of tested quantum particles in
 step~\ref{Ent:error-rate_test}.
 And in this case, they share a copy of the EPR-like state
 $|\Phi_{[a],0,0}\rangle$ per qudit transfer just after step~\ref{Ent:measure}.
 A simple-minded way to understand the origin of security of this scheme is
 that as Alice puts each shared EPR-like state in a Hilbert subspace, which is
 not known to Eve, she has a non-negligible chance of disturbing the signal if
 she guesses this subspace incorrectly.

\subsection{Reduction To Two Prepare-And-Measure-Based Schemes Known As
 Schemes~B And~C}
\label{Subsec:PMQKD}
 Consider the unitary operation $\text{BADD}$ for Alice and Bob to separately
 add their first qudit to their second qudit in the computational basis.
 Clearly,
\begin{align}
 & \text{BADD}(|\Phi_{[a],b,\bar{c}}\rangle \otimes |\Phi_{[a'],b',\bar{c}'}
  \rangle) \nonumber \\
 ={} & |\Phi_{[a],b,\bar{c}-\bar{c}'}\rangle \otimes |\Phi_{[a]+[a'],b+b',
  \bar{c}'}\rangle .
 \label{E:BADD_action}
\end{align}
 Consider also the unitary operation that acts on the computational basis
 according to
\begin{equation}
 H(|b\rangle) =
 \begin{cases}
  (|b\rangle + |b+1\rangle)/\sqrt{2} & \text{ if } b\in GF(N)/GF(2) , \\
  (-|b\rangle + |b+1\rangle)/\sqrt{2} & \text{ otherwise.}
 \end{cases}
 \label{E:extended_Had_def}
\end{equation}
 Then
\begin{align}
 H\otimes H (|\Phi_{[a],b,\bar{c}}\rangle) &= (-1)^{\bar{b}_0} |\Phi_{[a],b+
  \bar{b}_0+\bar{c},\bar{b}_0}\rangle \nonumber \\
 &= (-1)^{\bar{b}_0} |\Phi_{[a],[b]+\bar{c},\bar{b}_0}\rangle
 \label{E:extended_Had_action}
\end{align}
 up to a global phase, where $\bar{b}_0 \in GF(2)$ denotes the constant term of
 the degree-$(n\mathord{-}1)$ polynomial expression for $b$ in $GF(2)[x]$.
 Clearly, both $H\otimes H$ and $\text{BADD}$ map basis states in
 ${\mathcal B}$ to itself up to an overall phase.
 Since the error-correction and privacy amplification procedure using the
 specially designed \CSS code in step~\ref{Ent:priv_ampl} of \SchemeA involves
 $H\otimes H$, $\text{BADD}$, standard basis measurement plus local quantum
 operation by Bob only, therefore Alice may push her final measurement forward
 in time.
 By the Shor-Preskill argument~\cite{SP00,Chau05}, we obtain an equally secure
 \PMQKD scheme that we called \SchemeB.
 
 To find the corresponding channel error estimation method for this equally
 secure \SchemeB, we consider the linear operators~\cite{Rains99,AK01}
\begin{equation}
 \mathtt{X}_u |i\rangle = |i+u\rangle \text{~and~} \ \mathtt{Z}_u |i\rangle =
 (-1)^{\Tr(u i)} |i\rangle
 \label{E:X_Z_L_def}
\end{equation}
 for all $u\in GF(N)$, where $\Tr(i) = i+i^2+i^4+\dots +i^{N/2}$ is the
 absolute trace of $i$.
 Then
\begin{equation}
 L_\lambda^{-1} \mathtt{X}_u \mathtt{Z}_v L_\lambda = \mathtt{X}_{\lambda^{-1}
 u} \mathtt{Z}_{\lambda v} .
 \label{E:conjugation}
\end{equation}
 Recall that in \SchemeA, Alice first prepares the state
 $|\Phi_{[a],0,0}\rangle$.
 Consider those shared states to which Alice and Bob have applied the
 operations $I\otimes L_\lambda$ and $I\otimes L_\lambda^{-1}$, respectively.
 Suppose Alice and Bob separately measure these shared states after passing
 through the insecure channel in the $B_1$ basis.
 Suppose further that Alice informs Bob of her measurement outcomes.
 Then from Eqs.~\eqref{E:basis_def}
 and~\eqref{E:extended_Had_action}--\eqref{E:conjugation}, Bob could deduce
 $[\lambda^{-1}u]+ \Tr(\lambda v)$ and hence both $[\lambda^{-1}u]$ and $\Tr
 (\lambda v)$ as these two variables are linearly independent over $GF(2)$.
 Since the solution of the equation $[\lambda^{-1}u] = [b]$ is $u = \lambda
 [b]$ or $u = \lambda([b]+1)$, the outcomes of the above measurement by Alice
 and Bob give estimates of $\tilde{e}_{\lambda [b],\bar{c}} +
 \tilde{e}_{\lambda ([b]+1),\bar{c}}$ for all $\lambda\in GF(N)^*$, $[b]\in
 GF(N)/GF(2)$ and $\bar{c}\in GF(2)$.
 Hence, our \EDQKD \SchemeA can be reduced to the following equally secure
 \PMQKD \SchemeB.

\par\medskip
{\bf The Modified \PMQKD Scheme (\SchemeB).}
\begin{enumerate}
 \item Alice randomly picks $\lambda\in GF(N)^*$ and prepares one of the basis
  states in $B_\lambda$ by randomly selecting the parameters $[a]$ and
  $\bar{c}$.
  He sends the state to Bob.
  \label{PM:prepare}
 \item Upon reception, Bob randomly picks $\lambda'\in GF(N)^*$ and measures
  his received state in the $B_{\lambda'}$ basis.
  \label{PM:measure}
 \item They publicly announce the values of $\lambda,\lambda'$ used and keep
  their state only if $\lambda=\lambda'$.
  They add the parameters $([a],\bar{c})$ describing their prepared and
  measured states to their raw key string.
  They repeat steps~\ref{PM:prepare}--\ref{PM:basis_reveal} until they have a
  sufficiently long raw key.
  \label{PM:basis_reveal}
 \item They estimate the values of $\tilde{e}_{\lambda [b],\bar{c}} +
  \tilde{e}_{\lambda ([b]+1),\bar{c}}$ by revealing (and discarding) a random
  sample of dits from the raw key to which Alice and Bob have applied
  $I\otimes L_\lambda$ and $I\otimes L_\lambda^{-1}$, respectively.
  They proceed only if the error rate is sufficiently small.
  \label{PM:error-rate_test}
 \item Alice and Bob apply classical error correction and privacy amplification
  to their remaining raw keys based on the classical $N$-ary code $C_1$ and
  classical binary code $C_2$ obeying $\{0\} \subset C_2\subset C_1$.
  Moreover, $C_1$ may be picked depending on the error syndrome of $C_2$ just
  like the privacy amplification procedure reported in Ref.~\cite{L01}.
  Specifically, we denote the $(N/2)$-ary vector formed by the $[a]$'s and the
  binary vector formed by the $\bar{c}$'s in Alice's remaining raw key by
  $\vec{a}$ and $\vec{c}$, respectively.
  Alice announces the error syndromes for $C_1$ of $\vec{a}$ and $C_2$ of
  $\vec{c}$.
  Bob subtracts them from his corresponding measured error syndromes and then
  uses the subtracted results to perform classical error corrections using
  codes $C_1$ ($C_2$) on his remaining raw key $\vec{a}'$ ($\vec{c}'$).
  For a sufficiently low noise level, the Bob's raw key after error correction
  should agree with Alice's.
  They now use the cosets $\vec{a}+C_1$ and $\vec{c}+C_2$ as their shared final
  key.
  \label{PM:priv_ampl}
\end{enumerate}

 Note that \SchemeB is analogous to the \Chauonefive scheme in
 Ref.~\cite{Chau15}.
 The most notable difference is that unlike the \Chauonefive scheme, the
 two-dimensional Hilbert subspaces used in state preparation and measurement
 are not revealed in \SchemeB.

 Since each element in $B_\lambda$ can be rewritten in the form $(|i\rangle\pm
 |j\rangle)/\sqrt{2}$ for some $i\ne j\in GF(N)$, the state preparation of
 \SchemeB in step~\ref{PM:prepare} is exactly the same as that of the
 \Chauonefive.
 While the state measurement procedure of \SchemeB in step~\ref{PM:measure} is
 a complete measurement and is different from the incomplete measurement used
 in the \Chauonefive, we could further change this step in \SchemeB to
 step~\ref{PM:measure}' below so that the hardware setup is identical to that
 of the \Chauonefive scheme.
 We call this further modified protocol \SchemeC.
\par\medskip
{\bf The Further Modified \PMQKD Scheme (\SchemeC).}
\begin{enumerate}
 \setcounter{enumi}{2}
 \vspace{-0.5em}
 \item[\theenumi'.] Upon reception, Bob randomly picks $\lambda'\in GF(N)^*$,
  $[a']\in GF(N)/GF(2)$ and measures his received state along
  $\{|\lambda' [a']\rangle\pm|\lambda'(1+[a'])\rangle\}/\sqrt{2}$.
  (Clearly, this is equivalent to randomly picking $i'\ne j'\in GF(N)$ and
  measuring the received state along $(|i'\rangle\pm|j'\rangle)/\sqrt{2}$ as in
  the measurement step in the \Chauonefive scheme.)
  Bob informs Alice to ignore her parameters $\lambda$, $[a]$ and $\bar{c}$ and
  repeat her state preparation and sending procedures in step~\ref{PM:prepare}
  of \SchemeB in case his measurement fails.
\end{enumerate}
 Note that Schemes~B and~C are equally secure.
 The reason is that Eve's action on the qudits cannot depend on the values of
 $\lambda'$'s and $[a]$'s used for she has no knowledge of them when the qudits
 pass through the insecure quantum channel.
 Consequently, the error rates $\tilde{e}_{b\bar{c}}$ experienced by the
 $|\Phi_{[a],0,0}\rangle$'s in the corresponding \EDQKD \SchemeA for those
 discarded and undiscarded qudits are the same.
 In summary, this further modification in \SchemeC allows easier experimental
 implementation than \SchemeB because complete measurement in the
 $B_{\lambda'}$ basis even for $N=4$ is not trivial.
 However, the key rate of \SchemeC will be lower than that of \SchemeB since
 more signals have to be discarded in step~\ref{PM:measure}'.
 We shall get back to this point in Sec.~\ref{Subsec:KeyRate} below.

 Finally, we remark that it is possible to apply two-way error correction and
 privacy amplification in Schemes~A--C similar to the one used in the
 \Chauonefive scheme~\cite{Chau15}.
 In fact, the conclusions on the error-tolerable capability of the \Chauonefive
 scheme using two-way entanglement purification in Ref.~\cite{Chau15} is
 directly applicable to our three schemes.
 In what follows, however, we focus on the performance of Schemes~A--C using
 the more practical one-way entanglement purification procedure~\cite{SP00},
 which gives a higher key rate when the channel noise is low at the expense of
 having a lower error-tolerable rate.

\section{Security And Performance Analysis}
\label{Sec:Security}

\subsection{The Unconditional Security Proof Of \SchemeA}
 Recall that in \SchemeA, Eve sees the same completely mixed density matrix for
 the quantum state that Alice sends to Bob in step~\ref{Ent:prepare}
 irrespective of the value of $\lambda$ used.
 So the quantum operation $\rho \mapsto {\mathcal E}(\rho) = \sum_i K_i \rho
 K_i^\dag$ Eve applies to the insecure quantum channel is independent of
 $\lambda$, where each Kraus operator used can be written as $K_i = \sum_{u,v
 \in GF(N)} g_{iuv} {\mathtt X}_u {\mathtt Z}_v$ for some $g_{iuv}\in
 {\mathbb C}$.
 Since $\sum_i K_i^\dag K_i = I$ and ${\mathtt Z}_v {\mathtt X}_u = (-1)^{\Tr
 (u v)} {\mathtt X}_u {\mathtt Z}_v$~\cite{AK01,Chau05}, we have $\sum_{i,u,v}
 |g_{iuv}|^2 = 1$ and $\sum_{i,u,v} g_{iuv}^* g_{i,u,v+w} = 0$ for all $w\ne
 0$.
 Consequently,
\begin{align}
 \tilde{e}_{b,\bar{c}} &= \langle \Phi_{[a],b,\bar{c}} |
  {\mathcal E}(|\Phi_{[a],0,0}\rangle\langle\Phi_{[a],0,0}|)
  |\Phi_{[a],b,\bar{c}}\rangle \nonumber \\
 &= \sum_i \sum'_{v,v'} g_{i b v}^* g_{i b v'} =
  \sum_i \sum'_v |g_{i b v}|^2 \equiv \sum'_v e_{bv} ,
 \label{E:e_def}
\end{align}
 where the primed sum is over those variables $v$ and/or $v'\in GF(N)$
 satisfying $\Tr(v) = \Tr(v') = \bar{c}$.
 Note that $I\otimes {\mathtt X}_u {\mathtt Z}_v |\Phi_{[a],b,\bar{c}}\rangle =
 I \otimes {\mathtt X}_{u'} {\mathtt Z}_{v'} |\Phi_{[a],b,\bar{c}}\rangle$ up
 to an irrelevant phase whenever $u=u'$ and $\Tr(v) = \Tr(v')$.
 Combined with Eq.~\eqref{E:e_def}, we conclude that Eve's attack through
 ${\mathcal E}$ is equivalent to the quantum operation $\rho \mapsto \sum_{u,v}
 e_{uv} {\mathtt X}_u {\mathtt Z}_v \rho \left( {\mathtt X}_u {\mathtt Z}_v
 \right)^\dag$.
 In this regard, we may interpret $e_{uv}$ as the probability that the qudit
 has experienced ${\mathtt X}_u {\mathtt Z}_v$ in the insecure quantum channel.

 Recall that we obtain estimates of $\tilde{e}_{\lambda [b],\bar{c}} +
 \tilde{e}_{\lambda ([b]+1),\bar{c}}$ for $[b]\in GF(N)/GF(2)$, $\lambda\in
 GF(N)^*$ and $\bar{c}\in GF(2)$ in step~\ref{Ent:error-rate_test} of \SchemeA.
 In the infinite key length limit, these estimates are exact.
 More importantly, $e_{uv}$'s must be consistent with these estimates through
 Eq.~\eqref{E:e_def}.
 The dimension of each qudit received by Bob is $N = 2^n$.
 So quantum Gilbert-Varshamov bound~\cite{CS96,Steane96} tells us that the \CSS
 code needed to perform the entanglement distillation in
 step~\ref{Ent:priv_ampl} of \SchemeA exists provided that~\cite{L01}
\begin{align}
 K &= n - \max h_2(\{ e_{uv} \}_{u,v\in GF(N)}) \nonumber \\
 &\equiv n + \max \sum_{u,v\in GF(N)} e_{uv} \log_2 e_{uv} > 0 ,
 \label{E:GV_bound}
\end{align}
 where the maximum is over all $e_{uv}$'s in $[0,1]$ that are consistent with
 the error rate estimates, namely, $\tilde{e}_{\lambda[b],\bar{c}} +
 \tilde{e}_{\lambda([b]+1),\bar{c}}$ for $[b]\in GF(N)/GF(2)$, $\lambda \in
 GF(N)^*$ and $\bar{c} \in GF(2)$.
 Once this (random) \CSS code exists, Alice and Bob can almost surely distill
 out almost perfect states each in the form $|\Phi_{[a],0,0}\rangle$.

 There are a few ways to define the key rate for a \QKD protocol.
 Here we extend the one used for qubit transfer in Ref.~\cite{RMP09}, which is
 experimentally meaningful, to qudit transfer by defining the secret key rate
 as the number of provably secure dits distilled divided by the number of qudit
 transferred in the limit of an arbitrary large number of qudit
 transfer.
 In the case of a lossless channel and perfect detectors, the secret key rates
 for \SchemeA and hence also \SchemeB equal
\begin{subequations}
\label{E:key_rates}
\begin{equation}
 R_\text{A} = R_\text{B} = \max (0,K/\{ n (N-1) \}) .
 \label{E:key_rate_schemes_AB}
\end{equation}
 Note that in the above expression, the $1/(N-1)$ factor is the probability
 that Alice's $\lambda$ agrees with Bob's $\lambda'$; and the $1/n$ factor
 converts the number of secret bits to dits.
 To summarize, both Schemes~A and~B can distill out a secret key provided that
 Eq.~\eqref{E:GV_bound} holds for all $e_{uv}$'s given that they obey the
 constraints coming from the measurement statistics in
 step~\ref{Ent:error-rate_test} of \SchemeA.
 Clearly, the resultant secret key is composable~\cite{Ben-Or05,RK05}.
 This completes our proof of the unconditional security.

 Finally, we remark that for \SchemeC, Alice and Bob will add $n$~bits to their
 raw keys if $\lambda=\lambda'$ and Bob's measurement in
 step~\ref{PM:measure}' is successful.
 As a result, the secret key rate for \SchemeC equals
\begin{equation}
 R_\text{C} = 2 R_\text{B} / N = \max(0,2K/\{n N (N-1)\}) .
 \label{E:key_rate_scheme_C}
\end{equation}
\end{subequations}
 Here, the extra factor of $2/N$ is the probability of having a successful
 measurement in step~\ref{PM:measure}'.

\begin{figure}[t]
 \hspace*{-1.85cm}\includegraphics[scale=0.4]{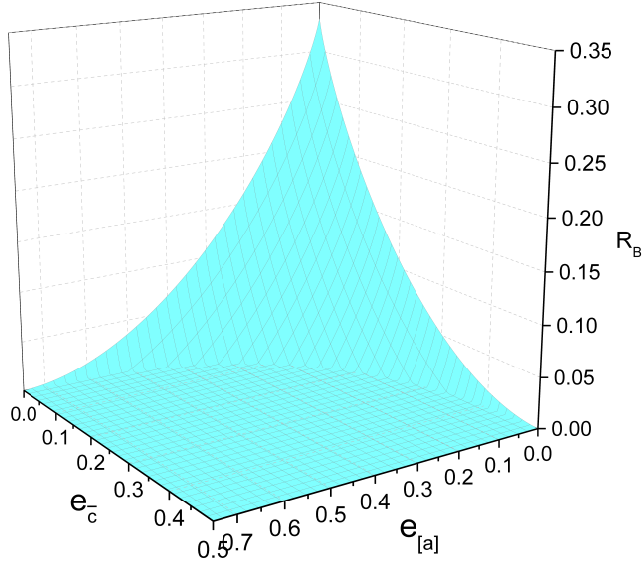}
 \caption{(color online) The secret key rate $R_\text{B}$ of \SchemeB as a
  function of $e_{[a]}$ and $e_{\bar{c}}$ for $N=4$ using one-way entanglement
  distillation.
  \label{F:key_rate_e_c_e_a_for_n=2}}
\end{figure}

\subsection{Key Rate Formulae For {S}chemes~A--C}
\label{Subsec:KeyRate}
 One may study the performance of Schemes~A--C using all the parameters
 obtained from step~\ref{Ent:error-rate_test} to constrain $e_{uv}$'s.
 But this approach is not very fruitful.
 Since the secret key comes from both $[a]$ and $\bar{c}$, it makes sense to
 gauge the performance of our \QKD scheme by one of the following two sets of
 parameters.
 The first set is the average bit error rate $e_{\bar{c}}$ of the $\bar{c}$'s
 and the average dit error rate $e_{[a]}$ of the $[a]$'s in the raw key.
 (Note that both $e_{\bar{c}}$ and $e_{[a]}$ are averaged over $\lambda$.)
 The second set is simply the bit error rate of the raw bit key string
 $e_\text{raw}$.

 Note that permuting the non-zero $u$ and $v$ indices in $e_{uv}$'s does not
 change the values $e_{\bar{c}}$, $e_{[a]}$ and $e_\text{raw}$.
 Combined with the convexity of $h_2$, we conclude that the maximum in the
 R.H.S. of Eq.~\eqref{E:GV_bound} is reached only if $e_{\mu 0} = e_{\mu'0}$,
 $e_{0\nu} = e_{0\nu'}$ and $e_{\mu\nu} = e_{\mu'\nu'}$ for all $\mu,\mu',\nu,
 \nu'\in GF(N)^*$.
 This observation greatly simplifies the computation of $K$ in
 Eq.~\eqref{E:GV_bound} as it becomes the easily manageable optimization
 problem involving four unknowns, namely, $A\equiv e_{00}$, $B\equiv 
 \sum_{\nu\in GF(N)^*} e_{0\nu} / (N-1)$, $C\equiv\sum_{\mu\in GF(N)^*}
 e_{\mu 0} / (N-1)$ and $D\equiv \sum_{\mu,\nu\in GF(N)^*} e_{\mu\nu} /
 (N-1)^2$ under the constraints
\begin{subequations}
\label{E:opt_constraints}
\begin{align}
 0 &\le A,B,C,D \le 1 ,
  \label{E:opt_constraint_begin} \\
 1 &= A+(N-1)(B+C)+(N-1)^2 D ,
  \label{E:opt_constraint_sum} \\
 e_{\bar{c}} &= \frac{N}{2} \{ B + (N-1)D \} , \\
 e_{[a]} &= (N-1) \{ C + (N-1)D \}
  \label{E:opt_constraint_end}
\end{align}
 for the case of finding $R_\text{B}(e_{\bar{c}},e_{[a]})$.
 And by putting in the additional constraint
\begin{equation}
 e_\text{raw} = \frac{1}{n} \left\{ e_{\bar{c}} + \frac{(n-1)N e_{[a]}}{(N-1)
 (N-2)} \right\} ,
 \label{E:e_raw}
\end{equation}
\end{subequations}
 one could determine $R_\text{B}(e_\text{raw})$.
 Note that the first term in the curly bracket in the R.H.S. of
 Eq.~\eqref{E:e_raw} comes from the fact that $1/n$ of the raw bits originates
 from the values of $\bar{c}$'s.
 For the second term, $\sum_{v\in GF(N)} e_{\mu v} = e_{[a]}/(N-1)$ for all
 $\mu\in GF(N)^*$ so that each type of dit error in $[a]$ occurs at a rate of
 $2 e_{[a]}/(N-1)$.
 Converting the dit $[a]$ to $(n-1)$~bits, the corresponding bit error rate
 becomes $2 e_{[a]}/(N-1) \times (N/4)/(N/2-1)$.
 Hence, the second term in Eq.~\eqref{E:e_raw} corresponds to the contribution
 of bit error rate in the value of $[a]$.

\begin{figure}[t]
 \hspace*{-0.6cm}\includegraphics[scale=0.39]{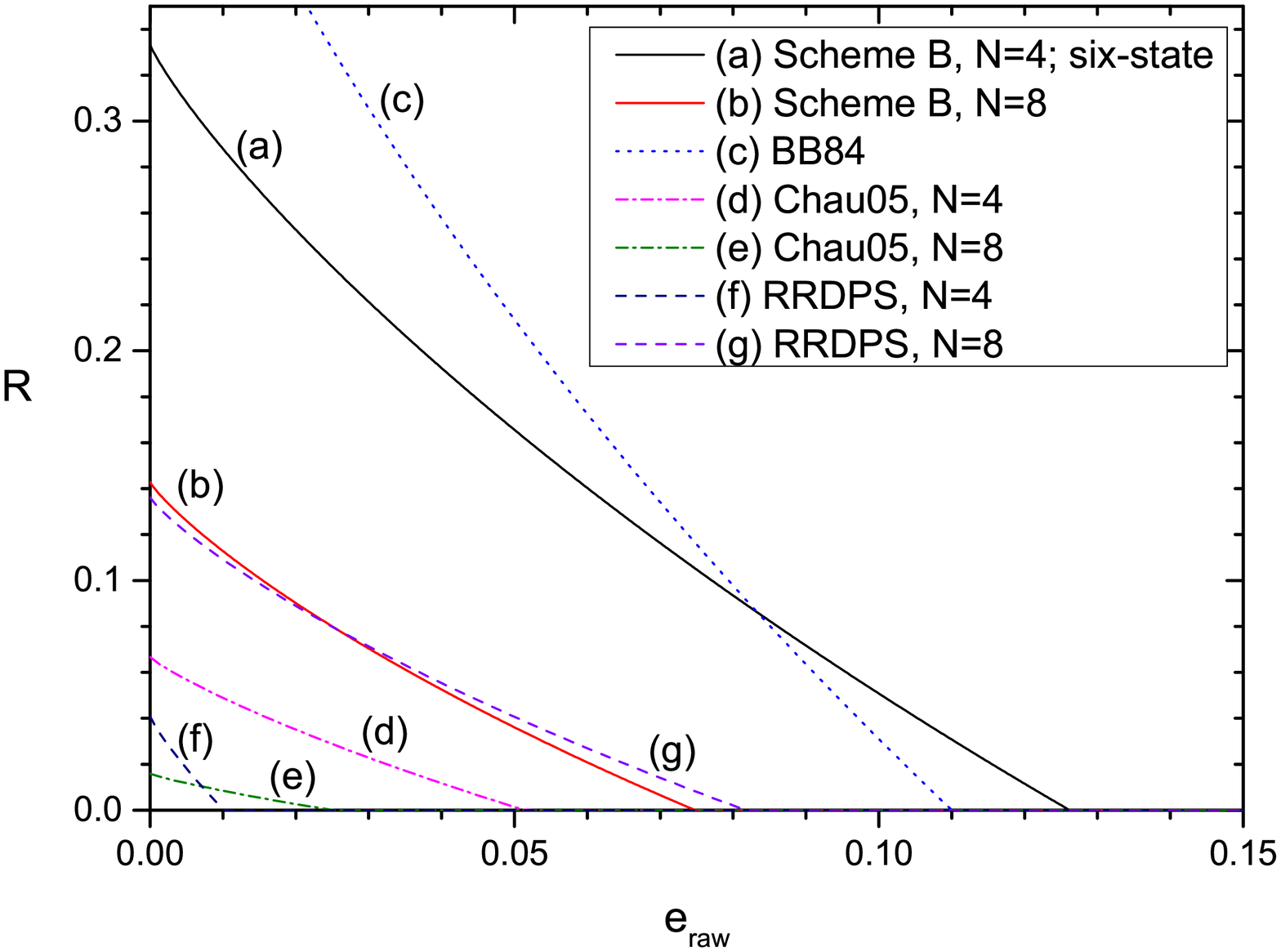}
 \caption{(color online) The secret key rates $R$ as a function of the bit
  error rate of the raw bit string $e_\text{raw}$ for various \PMQKD schemes in
  the one-way communication setting.
  \SchemeB, \BB, \Chauzerofive and \RRDPS are shown in solid, dot, dash-dot and
  dash curves, respectively.
  Note that the six-state scheme here uses non-degenerate quantum code for
  entanglement distillation.
  \label{F:key_rate_e_raw_for_n=2}}
\end{figure}

 Fig.~\ref{F:key_rate_e_c_e_a_for_n=2} plots the secret key rate $R_\text{B}$
 of \SchemeB for $N=4$ using one-way classical communication for fixed
 $e_{\bar{c}}$ and $e_{[a]}$ by numerically optimizing Eqs.~\eqref{E:GV_bound}
 and~\eqref{E:key_rate_schemes_AB} subject to the constraints in
 Eqs.~\eqref{E:opt_constraint_begin}--\eqref{E:opt_constraint_end}.
 It shows that the maximum tolerable error rate $e_{\bar{c}}$ ($e_{[a]}$) can
 be very high when $e_{[a]}$ ($e_{\bar{c}}$) is low.
 So, when $e_{[a]}$ is small, Alice and Bob may drop all the raw bits
 originated from $[a]$'s and use only the raw bits originated from $\bar{c}$'s
 to distill the secret key similar to the method used in Ref.~\cite{Chau15}.
 In this way, the maximum tolerable $e_{\bar{c}}$ can be as large as $1/2$ at
 the expense of having a very low secret key rate.
 Similar conclusion is drawn when $e_{\bar{c}}$ is small as Alice and Bob may
 keep only those raw bits generated from the measurement of $[a]$'s.
 However, we shall not pursue further along this direction here.

\begin{table}[t]
\begin{tabular}{||c|c|c|c||}
 \hline\hline
 Scheme & $N$\footnote{The Hilbert space dimension of the quantum information
   carrier.}
  & $R(0)$\footnote{The secret key rate in the noiseless and lossless
   situation.}
  & $e_\text{raw}^{\max}$\footnote{The maximum provably secure bit error rate
   of the raw key.} \\
 \hline
 \BB~\cite{BB84,SP00} & $2$ & $1/2$ & $11.0\%$ \\
 \hline
 Six-state~\cite{Bruss98,L01} & $2$ & $1/3$ & $12.7\%$ \\
 \hline
 \Chauzerofive~\cite{Chau05} & $4$ & $1/15$ & $5.1\%$ \\
 \cline{2-4}
 & $8$ & $1/63$ & $2.5\%$ \\
 \hline
 \RRDPS~\cite{SYK14} & $4$ & $0.041$ & $1.0\%$ \\
 \cline{2-4}
 & $8$ & $0.136$ & $8.2\%$ \\
 \hline
 \SchemeB & $4$ & $1/3$ & $12.6\%$ \\
 \cline{2-4}
 & $8$ & $1/7$ & $7.5\%$ \\
 \hline
 \SchemeC & $4$ & $1/6$ & $12.6\%$ \\
 \cline{2-4}
 & $8$ & $1/28$ & $7.5\%$ \\
 \hline\hline
\end{tabular}
 \caption{Performance of various \PMQKD schemes using one-way entanglement
  distillation.
  Here, the $e_\text{raw}^{\max}$ for the six-state scheme is for the case of
  using degenerate quantum code to perform entanglement distillation.
  \label{T:compare}}
\end{table}

\subsection{Comparison With Other Provably Secure \PMQKD Schemes}
\label{Subsec:Comparison}
 We now compare the performance of \SchemeB with a few well-known qubit- and
 qudit-based \PMQKD schemes.
 And we focus our comparison for the case when the quantum channel is lossless
 and that the detectors are prefect without dark counts and dead time in the
 limit of an infinitely long raw key.
 (We shall briefly discuss photon insertion loss in the experimental apparatus
 toward the end of this subsection.)
 In addition, we only consider the case of one-way privacy amplification.
 For qubit-based schemes, we choose the \BB~\cite{BB84} and the six-state
 schemes~\cite{Bruss98}.
 Note that for qudit-based \PMQKD schemes, unconditional security proofs are
 known only for a few of them.
 The representative examples are the \Chauzerofive scheme~\cite{Chau05} and the
 \RRDPS scheme~\cite{SYK14}.
 That is why we choose these two in our comparison.

 Fig.~\ref{F:key_rate_e_raw_for_n=2} depicts the secret key rates $R$ for
 various \PMQKD schemes using one-way entanglement distillation with
 non-degenerate quantum code as a function of $e_\text{raw}$.
 And Table~\ref{T:compare} summarizes the values of $R$ when $e_\text{raw}=0$
 as well as the maximum provably secure bit error rate of the raw key for these
 schemes.
 For the cases of $N=4$ and $8$ in \SchemeB, the secret key rates are computed
 by a similar numerical optimization procedure used to obtain
 Fig.~\ref{F:key_rate_e_c_e_a_for_n=2}.
 (For \SchemeC in both cases, the rate is one half of that for \SchemeB.)
 It shows that the maximum provably secure bit error rate $e_\text{raw}$
 decreases as $N$ increases.

 The most eye-catching feature in Fig.~\ref{F:key_rate_e_raw_for_n=2} is that
 the secret key rates of the six-state scheme~\cite{Bruss98} and \SchemeB for
 $N=4$ seem to agree.
 (That is to say, for $N=4$, \SchemeB and hence \SchemeC can tolerate up to
 12.6\% bit error rate~\cite{L01}.)
 Here we prove that it is indeed the case.

 First, there is a unique property for $N=4$, namely, that the bit error rate
 of the raw key $e_\text{raw}$ in Eq.~\eqref{E:e_raw} is unaltered by swapping
 $e_{\mu 0}$ with $e_{0\mu}$ for all $\mu\in GF(N)^*$ although this swapping
 may change the values of $e_{\bar{c}}$ and $e_{[a]}$.
 By convexity of $h_2$, the value of $K$ in Eq.~\eqref{E:GV_bound} is minimized
 only if $B=C$.
 Thus, Eq.~\eqref{E:GV_bound} can be written as an extremization over a single
 variable $D$ after eliminating $A$ and $B$ via the constraints given by
 Eqs.~\eqref{E:opt_constraint_sum}--\eqref{E:e_raw}.
 By finding the turning point of the resultant expression as a function of $D$,
 we conclude that $K$ is minimized when $D= e_\text{raw}^2 / 4$ [and hence
 $B=C=e_\text{raw}/2 - 3e_\text{raw}^2/4 = (e_\text{raw}/2)
 (1-3e_\text{raw}/2)$ and $A=1-3e_\text{raw}+9e_\text{raw}^2/4=(1-3e_\text{raw}
 /2)^2$].
 Upon simplification, we have
\begin{align}
 K &= 2+A\log_2 A+3B\log_2 B+3C\log_2 C+9D\log_2 D \nonumber \\
 &= 2 \left\{ 1 + \left( 1- \frac{3 e_\text{raw}}{2}
  \right) \log_2 \left( 1 - \frac{3 e_\text{raw}}{2} \right) \right. \nonumber
  \\
 & \qquad \quad \left. + \frac{3 e_\text{raw}}{2} \log_2 \left(
  \frac{e_\text{raw}}{2} \right) \right\} .
 \label{E:SchemeB_n=2_K}
\end{align}
 From Eq.~\eqref{E:key_rate_schemes_AB}, we conclude that the secret key rate
 of \SchemeB as a function of $e_\text{raw}$ is the same as that of the
 six-state scheme by one-way entanglement distillation using non-degenerate
 codes~\cite{L01}.
 We do not believe that this is coincidental for these two very different
 schemes to have the same secret key rate.
 But we have no idea why.

 The most error-tolerant \PMQKD scheme using $N$-dimensional qudits is the
 \Chauzerofive scheme, which can tolerate up to $e_\text{raw} = 35.6\%$ using
 two-way classical communication for $N=4$~\cite{Chau05}.
 To find the maximum tolerable error rate using one-way communication for the
 \Chauzerofive scheme, we use the fact that the scheme effectively depolarizes
 the quantum error so that $e_{uv} = e_{u' v'}$ for all $(u,v),(u',v') \ne
 (0,0)$.
 Hence, Eq.~\eqref{E:GV_bound} becomes
\begin{equation}
 K = n + e_{00} \log_2 e_{00} + (1-e_{00}) \log_2 \left( \frac{1-e_{00}}{N^2-1}
 \right) .
 \label{E:K_for_Chauzerofive}
\end{equation}
 When $N=4$, $K=0$ at $e_{00} = 0.710$.
 Thus, the \Chauzerofive scheme using one-way entanglement distillation can
 tolerate up to $(1-e_{00})N/(N^2-1)\times (N/2)/(N-1) = 5.1\%$ bit error rate
 for $N=4$.
 By the same analysis, the \Chauzerofive scheme using one-way entanglement
 distillation can tolerate up to 2.5\% bit error rate.
 
 For the \RRDPS scheme using $N$-dimensional qudits with one-way privacy
 amplification, the secret key rate is given by~\cite{SYK14}
\begin{equation}
 R_\text{RRDPS} = \frac{1}{\log_2 N} \left\{ 1 - h_2(\frac{1}{N-1}) -
 h_2(e_\text{raw}) \right\} ,
 \label{E:secret_key_rate_RRDPS}
\end{equation}
 where $h_2(e) = -e \log_2 e - (1-e) \log_2 (1-e)$.
 (Note that the extra $1/\log_2 N$ factor, which does not appear in
 Ref.~\cite{SYK14}, converts the number of bits to the number of dits in the
 raw key.)
 That is to say, it can tolerate up to a bit error rate of $1.0\%$ and $8.2\%$
 for $N=4$ and $8$, respectively.

 These findings are summarized in Fig.~\ref{F:key_rate_e_raw_for_n=2} and
 Table~\ref{T:compare}.
 They show that, for both $N=4$ and $8$, the error-tolerant capability of
 Schemes~B and~C using one-way classical communication is better than the
 \Chauzerofive~\cite{Chau05} and the \RRDPS~\cite{SYK14} schemes.
 Table~\ref{T:compare} also shows the secret key rate in the noiseless
 situation.
 In this situation, and for $N=4$, the secret key rate of \SchemeB is the same
 as that of the six-state scheme and is much higher than those of the
 \Chauzerofive~\cite{Chau05} and the \RRDPS~\cite{SYK14} schemes but lower than
 the \BB~\cite{BB84}.
 All in all, we conclude that in terms of the secret key rate in the noiseless
 limit and the maximum tolerable provably secure bit error rate of the raw key
 using one-way entanglement distillation, both Schemes~B and~C can be ranked
 among the best all-rounded \PMQKD schemes for $N=4$.
 Therefore, from Fig.~\ref{F:key_rate_e_raw_for_n=2}, when restricted to
 one-way entanglement distillation, the most economical way for Alice and Bob
 to share their secret key is to use the \BB in case the channel noise is low
 (when $e_\text{raw} \lesssim 9\%$) and either the six-state scheme or \SchemeB
 with $N=4$ if the channel noise is high (when $9\% \lesssim e_\text{raw}
 \lesssim 12\%$) provided that errors and losses in the labs of Alice and Bob
 are negligibly small.

 For actual experimental setup, we have already pointed out that the state
 preparation of Schemes~B and~C is the same as that of the \Chauonefive scheme
 in Ref.~\cite{Chau15}.
 For \SchemeB, the measurement is more complicated than in Ref.~\cite{Chau15}
 for the present scheme requires complete measurement.
 The measurement can either be done by active or passive basis selection, the
 latter case can be done by adapting the method used by Muller \emph{et al.} in
 Ref.~\cite{MBG93} using $(N-1)N$~photon detectors, which is barely feasible
 though not very economical for $N=4$ due to the large number of detectors
 required.
 Whereas for \SchemeC, the measurement can be done in exactly the same way as
 in the \Chauonefive scheme~\cite{Chau15}, which can be directly adapted from
 the measurement part of various \RRDPS
 experiments~\cite{RRDPS-Expt1,RRDPS-Expt2,RRDPS-Expt3}.
 It is instructive to carry out actual experiments using Schemes~B and~C and
 compare their performances with that of the six-state scheme; and we are
 going to do so.

\begin{acknowledgments}
 We would like to thank Tieqiao Huang and H.-K. Lo for their discussions.
\end{acknowledgments}

\bibliographystyle{apsrev4-1}
\bibliography{qc69.6}

\begin{thebibliography}{23}%
\makeatletter
\providecommand \@ifxundefined [1]{%
 \@ifx{#1\undefined}
}%
\providecommand \@ifnum [1]{%
 \ifnum #1\expandafter \@firstoftwo
 \else \expandafter \@secondoftwo
 \fi
}%
\providecommand \@ifx [1]{%
 \ifx #1\expandafter \@firstoftwo
 \else \expandafter \@secondoftwo
 \fi
}%
\providecommand \natexlab [1]{#1}%
\providecommand \enquote  [1]{``#1''}%
\providecommand \bibnamefont  [1]{#1}%
\providecommand \bibfnamefont [1]{#1}%
\providecommand \citenamefont [1]{#1}%
\providecommand \href@noop [0]{\@secondoftwo}%
\providecommand \href [0]{\begingroup \@sanitize@url \@href}%
\providecommand \@href[1]{\@@startlink{#1}\@@href}%
\providecommand \@@href[1]{\endgroup#1\@@endlink}%
\providecommand \@sanitize@url [0]{\catcode `\\12\catcode `\$12\catcode
  `\&12\catcode `\#12\catcode `\^12\catcode `\_12\catcode `\%12\relax}%
\providecommand \@@startlink[1]{}%
\providecommand \@@endlink[0]{}%
\providecommand \url  [0]{\begingroup\@sanitize@url \@url }%
\providecommand \@url [1]{\endgroup\@href {#1}{\urlprefix }}%
\providecommand \urlprefix  [0]{URL }%
\providecommand \Eprint [0]{\href }%
\providecommand \doibase [0]{http://dx.doi.org/}%
\providecommand \selectlanguage [0]{\@gobble}%
\providecommand \bibinfo  [0]{\@secondoftwo}%
\providecommand \bibfield  [0]{\@secondoftwo}%
\providecommand \translation [1]{[#1]}%
\providecommand \BibitemOpen [0]{}%
\providecommand \bibitemStop [0]{}%
\providecommand \bibitemNoStop [0]{.\EOS\space}%
\providecommand \EOS [0]{\spacefactor3000\relax}%
\providecommand \BibitemShut  [1]{\csname bibitem#1\endcsname}%
\let\auto@bib@innerbib\@empty
\bibitem [{\citenamefont {Chau}(2015)}]{Chau15}%
  \BibitemOpen
  \bibfield  {author} {\bibinfo {author} {\bibfnamefont {H.~F.}\ \bibnamefont
  {Chau}},\ }\href@noop {} {\bibfield  {journal} {\bibinfo  {journal} {Phys.
  Rev. A}\ }\textbf {\bibinfo {volume} {92}},\ \bibinfo {pages} {062324}
  (\bibinfo {year} {2015})}\BibitemShut {NoStop}%
\bibitem [{\citenamefont {Scarani}\ \emph {et~al.}(2009)\citenamefont
  {Scarani}, \citenamefont {Bechmann-Pasquinucci}, \citenamefont {Cerf},
  \citenamefont {Du{\v{s}}ek}, \citenamefont {L{\"u}tkenhaus},\ and\
  \citenamefont {Peev}}]{RMP09}%
  \BibitemOpen
  \bibfield  {author} {\bibinfo {author} {\bibfnamefont {V.}~\bibnamefont
  {Scarani}}, \bibinfo {author} {\bibfnamefont {H.}~\bibnamefont
  {Bechmann-Pasquinucci}}, \bibinfo {author} {\bibfnamefont {N.~J.}\
  \bibnamefont {Cerf}}, \bibinfo {author} {\bibfnamefont {M.}~\bibnamefont
  {Du{\v{s}}ek}}, \bibinfo {author} {\bibfnamefont {N.}~\bibnamefont
  {L{\"u}tkenhaus}}, \ and\ \bibinfo {author} {\bibfnamefont {M.}~\bibnamefont
  {Peev}},\ }\href@noop {} {\bibfield  {journal} {\bibinfo  {journal} {Rev.
  Mod. Phys.}\ }\textbf {\bibinfo {volume} {81}},\ \bibinfo {pages} {1301}
  (\bibinfo {year} {2009})}\BibitemShut {NoStop}%
\bibitem [{\citenamefont {Bennett}\ and\ \citenamefont
  {Brassard}(1984)}]{BB84}%
  \BibitemOpen
  \bibfield  {author} {\bibinfo {author} {\bibfnamefont {C.~H.}\ \bibnamefont
  {Bennett}}\ and\ \bibinfo {author} {\bibfnamefont {G.}~\bibnamefont
  {Brassard}},\ }in\ \href@noop {} {\emph {\bibinfo {booktitle} {Proc. IEEE
  Int. Conf. on Computers, Systems and Signal Processing}}}\ (\bibinfo
  {publisher} {IEEE Press},\ \bibinfo {year} {1984})\ pp.\ \bibinfo {pages}
  {175--179},\ \bibinfo {note} {reprinted with corrections in \emph{Theo. Comp.
  Sci.} \textbf{560}, 7 (2014)}\BibitemShut {NoStop}%
\bibitem [{\citenamefont {Bechmann-Pasquinucci}\ and\ \citenamefont
  {Peres}(2000)}]{BP00}%
  \BibitemOpen
  \bibfield  {author} {\bibinfo {author} {\bibfnamefont {H.}~\bibnamefont
  {Bechmann-Pasquinucci}}\ and\ \bibinfo {author} {\bibfnamefont
  {A.}~\bibnamefont {Peres}},\ }\href@noop {} {\bibfield  {journal} {\bibinfo
  {journal} {Phys. Rev. Lett.}\ }\textbf {\bibinfo {volume} {85}},\ \bibinfo
  {pages} {3313} (\bibinfo {year} {2000})}\BibitemShut {NoStop}%
\bibitem [{\citenamefont {Bechmann-Pasquinucci}\ and\ \citenamefont
  {Tittel}(2000)}]{BT00}%
  \BibitemOpen
  \bibfield  {author} {\bibinfo {author} {\bibfnamefont {H.}~\bibnamefont
  {Bechmann-Pasquinucci}}\ and\ \bibinfo {author} {\bibfnamefont
  {W.}~\bibnamefont {Tittel}},\ }\href@noop {} {\bibfield  {journal} {\bibinfo
  {journal} {Phys. Rev. A}\ }\textbf {\bibinfo {volume} {61}},\ \bibinfo
  {pages} {062308} (\bibinfo {year} {2000})}\BibitemShut {NoStop}%
\bibitem [{\citenamefont {Cerf}\ \emph {et~al.}(2002)\citenamefont {Cerf},
  \citenamefont {Bourennane}, \citenamefont {Karlsson},\ and\ \citenamefont
  {Gisin}}]{CBKG02}%
  \BibitemOpen
  \bibfield  {author} {\bibinfo {author} {\bibfnamefont {N.~J.}\ \bibnamefont
  {Cerf}}, \bibinfo {author} {\bibfnamefont {M.}~\bibnamefont {Bourennane}},
  \bibinfo {author} {\bibfnamefont {A.}~\bibnamefont {Karlsson}}, \ and\
  \bibinfo {author} {\bibfnamefont {N.}~\bibnamefont {Gisin}},\ }\href@noop {}
  {\bibfield  {journal} {\bibinfo  {journal} {Phys. Rev. Lett.}\ }\textbf
  {\bibinfo {volume} {88}},\ \bibinfo {pages} {127902} (\bibinfo {year}
  {2002})}\BibitemShut {NoStop}%
\bibitem [{\citenamefont {Chau}(2005)}]{Chau05}%
  \BibitemOpen
  \bibfield  {author} {\bibinfo {author} {\bibfnamefont {H.~F.}\ \bibnamefont
  {Chau}},\ }\href@noop {} {\bibfield  {journal} {\bibinfo  {journal} {IEEE
  Trans. Inf. Theo.}\ }\textbf {\bibinfo {volume} {51}},\ \bibinfo {pages}
  {1451} (\bibinfo {year} {2005})}\BibitemShut {NoStop}%
\bibitem [{\citenamefont {Sasaki}\ \emph {et~al.}(2014)\citenamefont {Sasaki},
  \citenamefont {Yamamoto},\ and\ \citenamefont {Koashi}}]{SYK14}%
  \BibitemOpen
  \bibfield  {author} {\bibinfo {author} {\bibfnamefont {T.}~\bibnamefont
  {Sasaki}}, \bibinfo {author} {\bibfnamefont {Y.}~\bibnamefont {Yamamoto}}, \
  and\ \bibinfo {author} {\bibfnamefont {M.}~\bibnamefont {Koashi}},\
  }\href@noop {} {\bibfield  {journal} {\bibinfo  {journal} {Nature}\ }\textbf
  {\bibinfo {volume} {509}},\ \bibinfo {pages} {475} (\bibinfo {year}
  {2014})}\BibitemShut {NoStop}%
\bibitem [{\citenamefont {Shor}\ and\ \citenamefont {Preskill}(2000)}]{SP00}%
  \BibitemOpen
  \bibfield  {author} {\bibinfo {author} {\bibfnamefont {P.~W.}\ \bibnamefont
  {Shor}}\ and\ \bibinfo {author} {\bibfnamefont {J.}~\bibnamefont
  {Preskill}},\ }\href@noop {} {\bibfield  {journal} {\bibinfo  {journal}
  {Phys. Rev. Lett.}\ }\textbf {\bibinfo {volume} {85}},\ \bibinfo {pages}
  {441} (\bibinfo {year} {2000})}\BibitemShut {NoStop}%
\bibitem [{\citenamefont {Bru{\ss}}(1998)}]{Bruss98}%
  \BibitemOpen
  \bibfield  {author} {\bibinfo {author} {\bibfnamefont {D.}~\bibnamefont
  {Bru{\ss}}},\ }\href@noop {} {\bibfield  {journal} {\bibinfo  {journal}
  {Phys. Rev. Lett.}\ }\textbf {\bibinfo {volume} {81}},\ \bibinfo {pages}
  {3018} (\bibinfo {year} {1998})}\BibitemShut {NoStop}%
\bibitem [{\citenamefont {Lin}\ and\ \citenamefont {{Costello,
  Jr.}}(2004)}]{FF}%
  \BibitemOpen
  \bibfield  {author} {\bibinfo {author} {\bibfnamefont {S.}~\bibnamefont
  {Lin}}\ and\ \bibinfo {author} {\bibfnamefont {D.~J.}\ \bibnamefont
  {{Costello, Jr.}}},\ }\href@noop {} {\emph {\bibinfo {title} {Error Control
  Coding}}},\ \bibinfo {edition} {2nd}\ ed.\ (\bibinfo  {publisher} {Prentice
  Hall},\ \bibinfo {address} {Upper Saddle River, NJ},\ \bibinfo {year}
  {2004})\ Chap.\ \bibinfo {chapter} {2.2--2.6}\BibitemShut {NoStop}%
\bibitem [{\citenamefont {Lo}(2001)}]{L01}%
  \BibitemOpen
  \bibfield  {author} {\bibinfo {author} {\bibfnamefont {H.-K.}\ \bibnamefont
  {Lo}},\ }\href@noop {} {\bibfield  {journal} {\bibinfo  {journal} {Quant.
  Inf. \& Comp.}\ }\textbf {\bibinfo {volume} {1}},\ \bibinfo {pages} {81}
  (\bibinfo {year} {2001})}\BibitemShut {NoStop}%
\bibitem [{\citenamefont {Gottesman}\ and\ \citenamefont {Lo}(2003)}]{GL03}%
  \BibitemOpen
  \bibfield  {author} {\bibinfo {author} {\bibfnamefont {D.}~\bibnamefont
  {Gottesman}}\ and\ \bibinfo {author} {\bibfnamefont {H.-K.}\ \bibnamefont
  {Lo}},\ }\href@noop {} {\bibfield  {journal} {\bibinfo  {journal} {IEEE
  Trans. Inf. Theo.}\ }\textbf {\bibinfo {volume} {49}},\ \bibinfo {pages}
  {457} (\bibinfo {year} {2003})}\BibitemShut {NoStop}%
\bibitem [{\citenamefont {Calderbank}\ and\ \citenamefont {Shor}(1996)}]{CS96}%
  \BibitemOpen
  \bibfield  {author} {\bibinfo {author} {\bibfnamefont {A.~R.}\ \bibnamefont
  {Calderbank}}\ and\ \bibinfo {author} {\bibfnamefont {P.~W.}\ \bibnamefont
  {Shor}},\ }\href@noop {} {\bibfield  {journal} {\bibinfo  {journal} {Phys.
  Rev. A}\ }\textbf {\bibinfo {volume} {54}},\ \bibinfo {pages} {1098}
  (\bibinfo {year} {1996})}\BibitemShut {NoStop}%
\bibitem [{\citenamefont {Steane}(1996)}]{Steane96}%
  \BibitemOpen
  \bibfield  {author} {\bibinfo {author} {\bibfnamefont {A.~M.}\ \bibnamefont
  {Steane}},\ }\href@noop {} {\bibfield  {journal} {\bibinfo  {journal} {Proc.
  Roy. Soc. Lond. A}\ }\textbf {\bibinfo {volume} {452}},\ \bibinfo {pages}
  {2551} (\bibinfo {year} {1996})}\BibitemShut {NoStop}%
\bibitem [{\citenamefont {Rains}(1999)}]{Rains99}%
  \BibitemOpen
  \bibfield  {author} {\bibinfo {author} {\bibfnamefont {E.~M.}\ \bibnamefont
  {Rains}},\ }\href@noop {} {\bibfield  {journal} {\bibinfo  {journal} {IEEE
  Trans. Inf. Theo.}\ }\textbf {\bibinfo {volume} {45}},\ \bibinfo {pages}
  {1827} (\bibinfo {year} {1999})}\BibitemShut {NoStop}%
\bibitem [{\citenamefont {Ashikhmin}\ and\ \citenamefont {Knill}(2001)}]{AK01}%
  \BibitemOpen
  \bibfield  {author} {\bibinfo {author} {\bibfnamefont {A.}~\bibnamefont
  {Ashikhmin}}\ and\ \bibinfo {author} {\bibfnamefont {E.}~\bibnamefont
  {Knill}},\ }\href@noop {} {\bibfield  {journal} {\bibinfo  {journal} {IEEE
  Trans. Inf. Theo.}\ }\textbf {\bibinfo {volume} {47}},\ \bibinfo {pages}
  {3065} (\bibinfo {year} {2001})}\BibitemShut {NoStop}%
\bibitem [{\citenamefont {Ben-Or}\ \emph {et~al.}(2005)\citenamefont {Ben-Or},
  \citenamefont {Horodecki}, \citenamefont {Leung}, \citenamefont {Mayers},\
  and\ \citenamefont {Oppenheim}}]{Ben-Or05}%
  \BibitemOpen
  \bibfield  {author} {\bibinfo {author} {\bibfnamefont {M.}~\bibnamefont
  {Ben-Or}}, \bibinfo {author} {\bibfnamefont {M.}~\bibnamefont {Horodecki}},
  \bibinfo {author} {\bibfnamefont {D.~W.}\ \bibnamefont {Leung}}, \bibinfo
  {author} {\bibfnamefont {D.}~\bibnamefont {Mayers}}, \ and\ \bibinfo {author}
  {\bibfnamefont {J.}~\bibnamefont {Oppenheim}},\ }in\ \href@noop {} {\emph
  {\bibinfo {booktitle} {Theory Of Cryptography: {S}econd Theory Of
  Cryptography Conference, TCC2005}}},\ \bibinfo {editor} {edited by\ \bibinfo
  {editor} {\bibfnamefont {J.}~\bibnamefont {Killian}}}\ (\bibinfo  {publisher}
  {Springer},\ \bibinfo {address} {Berlin},\ \bibinfo {year} {2005})\ pp.\
  \bibinfo {pages} {386--406}\BibitemShut {NoStop}%
\bibitem [{\citenamefont {Renner}\ and\ \citenamefont
  {K{\"o}nig}(2005)}]{RK05}%
  \BibitemOpen
  \bibfield  {author} {\bibinfo {author} {\bibfnamefont {R.}~\bibnamefont
  {Renner}}\ and\ \bibinfo {author} {\bibfnamefont {R.}~\bibnamefont
  {K{\"o}nig}},\ }in\ \href@noop {} {\emph {\bibinfo {booktitle} {Theory Of
  Cryptography: {S}econd Theory Of Cryptography Conference, TCC2005}}},\
  \bibinfo {editor} {edited by\ \bibinfo {editor} {\bibfnamefont
  {J.}~\bibnamefont {Killian}}}\ (\bibinfo  {publisher} {Springer},\ \bibinfo
  {address} {Berlin},\ \bibinfo {year} {2005})\ pp.\ \bibinfo {pages}
  {407--425}\BibitemShut {NoStop}%
\bibitem [{\citenamefont {Muller}\ \emph {et~al.}(1993)\citenamefont {Muller},
  \citenamefont {Breguet},\ and\ \citenamefont {Gisin}}]{MBG93}%
  \BibitemOpen
  \bibfield  {author} {\bibinfo {author} {\bibfnamefont {A.}~\bibnamefont
  {Muller}}, \bibinfo {author} {\bibfnamefont {J.}~\bibnamefont {Breguet}}, \
  and\ \bibinfo {author} {\bibfnamefont {N.}~\bibnamefont {Gisin}},\
  }\href@noop {} {\bibfield  {journal} {\bibinfo  {journal} {Europhys. Lett.}\
  }\textbf {\bibinfo {volume} {23}},\ \bibinfo {pages} {383} (\bibinfo {year}
  {1993})}\BibitemShut {NoStop}%
\bibitem [{\citenamefont {Takesue}\ \emph {et~al.}(2015)\citenamefont
  {Takesue}, \citenamefont {Sasaki}, \citenamefont {Tamaki},\ and\
  \citenamefont {Koashi}}]{RRDPS-Expt1}%
  \BibitemOpen
  \bibfield  {author} {\bibinfo {author} {\bibfnamefont {H.}~\bibnamefont
  {Takesue}}, \bibinfo {author} {\bibfnamefont {T.}~\bibnamefont {Sasaki}},
  \bibinfo {author} {\bibfnamefont {K.}~\bibnamefont {Tamaki}}, \ and\ \bibinfo
  {author} {\bibfnamefont {M.}~\bibnamefont {Koashi}},\ }\href@noop {}
  {\bibfield  {journal} {\bibinfo  {journal} {Nature Photonics}\ }\textbf
  {\bibinfo {volume} {9}},\ \bibinfo {pages} {827} (\bibinfo {year}
  {2015})}\BibitemShut {NoStop}%
\bibitem [{\citenamefont {Wang}\ \emph {et~al.}(2015)\citenamefont {Wang},
  \citenamefont {Yin}, \citenamefont {Chen}, \citenamefont {He}, \citenamefont
  {Song}, \citenamefont {Li}, \citenamefont {Zhang}, \citenamefont {Zhou},
  \citenamefont {Guo},\ and\ \citenamefont {Han}}]{RRDPS-Expt2}%
  \BibitemOpen
  \bibfield  {author} {\bibinfo {author} {\bibfnamefont {S.}~\bibnamefont
  {Wang}}, \bibinfo {author} {\bibfnamefont {Z.-Q.}\ \bibnamefont {Yin}},
  \bibinfo {author} {\bibfnamefont {W.}~\bibnamefont {Chen}}, \bibinfo {author}
  {\bibfnamefont {D.-Y.}\ \bibnamefont {He}}, \bibinfo {author} {\bibfnamefont
  {X.-T.}\ \bibnamefont {Song}}, \bibinfo {author} {\bibfnamefont {H.-W.}\
  \bibnamefont {Li}}, \bibinfo {author} {\bibfnamefont {L.-J.}\ \bibnamefont
  {Zhang}}, \bibinfo {author} {\bibfnamefont {Z.}~\bibnamefont {Zhou}},
  \bibinfo {author} {\bibfnamefont {G.-C.}\ \bibnamefont {Guo}}, \ and\
  \bibinfo {author} {\bibfnamefont {Z.-F.}\ \bibnamefont {Han}},\ }\href@noop
  {} {\bibfield  {journal} {\bibinfo  {journal} {Nature Photonics}\ }\textbf
  {\bibinfo {volume} {9}},\ \bibinfo {pages} {832} (\bibinfo {year}
  {2015})}\BibitemShut {NoStop}%
\bibitem [{\citenamefont {Li}\ \emph {et~al.}(2016)\citenamefont {Li},
  \citenamefont {Cao}, \citenamefont {Dai}, \citenamefont {Lin}, \citenamefont
  {Zhang}, \citenamefont {Chen}, \citenamefont {Xu}, \citenamefont {Guan},
  \citenamefont {Liao}, \citenamefont {Yin}, \citenamefont {Zhang},
  \citenamefont {Ma}, \citenamefont {Peng},\ and\ \citenamefont
  {Pan}}]{RRDPS-Expt3}%
  \BibitemOpen
  \bibfield  {author} {\bibinfo {author} {\bibfnamefont {Y.-H.}\ \bibnamefont
  {Li}}, \bibinfo {author} {\bibfnamefont {Y.}~\bibnamefont {Cao}}, \bibinfo
  {author} {\bibfnamefont {H.}~\bibnamefont {Dai}}, \bibinfo {author}
  {\bibfnamefont {J.}~\bibnamefont {Lin}}, \bibinfo {author} {\bibfnamefont
  {Z.}~\bibnamefont {Zhang}}, \bibinfo {author} {\bibfnamefont
  {W.}~\bibnamefont {Chen}}, \bibinfo {author} {\bibfnamefont {Y.}~\bibnamefont
  {Xu}}, \bibinfo {author} {\bibfnamefont {J.-Y.}\ \bibnamefont {Guan}},
  \bibinfo {author} {\bibfnamefont {S.-K.}\ \bibnamefont {Liao}}, \bibinfo
  {author} {\bibfnamefont {J.}~\bibnamefont {Yin}}, \bibinfo {author}
  {\bibfnamefont {Q.}~\bibnamefont {Zhang}}, \bibinfo {author} {\bibfnamefont
  {X.}~\bibnamefont {Ma}}, \bibinfo {author} {\bibfnamefont {C.-Z.}\
  \bibnamefont {Peng}}, \ and\ \bibinfo {author} {\bibfnamefont {J.-W.}\
  \bibnamefont {Pan}},\ }\href@noop {} {\bibfield  {journal} {\bibinfo
  {journal} {Phys. Rev. A}\ }\textbf {\bibinfo {volume} {93}},\ \bibinfo
  {pages} {030302(R)} (\bibinfo {year} {2016})}\BibitemShut {NoStop}%
\end{thebibliography}%
\end{document}